\documentclass[%
prl,
twocolumn,
superscriptaddress,
 amsmath,amssymb,
 aps,
]{revtex4-1}

\usepackage{graphicx}
\usepackage{dcolumn}
\usepackage{bm}

\begin{document}

\title{Spiro-Conjugated Molecular Junctions: between Jahn-Teller Distortion and Destructive Quantum Interference}
\author{Jakub K. Sowa}
\email{jakub.sowa@materials.ox.ac.uk}
\author{Jan A. Mol}
\author{G. Andrew D. Briggs}
\affiliation{
 Department of Materials, University of Oxford, Parks Road, Oxford OX1 3PH, United Kingdom
}

\author{Erik M. Gauger}
\affiliation{SUPA, Institute of Photonics and Quantum Sciences, Heriot-Watt University, EH14 4AS, United Kingdom}

\date{\today}
\begin{abstract}
The quest for molecular structures exhibiting strong quantum interference effects in the transport setting has long been on the forefront of chemical research.
Here, we establish theoretically that the unusual geometry of spiro-conjugated systems gives rise to complete destructive interference in the resonant-transport regime. This results in a current blockade of the type not present in meta-connected benzene or similar molecular structures.
We further show that these systems can undergo a transport-driven Jahn-Teller distortion which can lift the aforementioned destructive-interference effects. The overall transport characteristics is determined by the interplay between the two phenomena. 
Spiro-conjugated systems may therefore serve as a novel platform for investigations of quantum interference and vibronic effects in the charge transport setting. 
The potential to control quantum interference in these systems can also turn them into attractive components in designing functional molecular circuits.
\end{abstract}

\maketitle

\textit{Introduction.}---Much of the development in the field of single-molecule electronics has been driven by the possibility of exploiting quantum interference (QI) effects to construct smaller and more efficient electronic devices. Hitherto, investigations of QI effects in single-molecule junctions utilised predominantly planar ring structures \cite{walter2004quantum,ke2008quantum,cardamone2006controlling,arroyo2013signatures,pickup2015new}, and cross-conjugate molecular systems \cite{solomon2008quantum,guedon2012observation,solomon2008things,koole2015electric,lambert2015basic,papadopoulos2006control}. Additionally, the vast majority of theoretical studies focused on the off-resonant transport regime, although QI effects in the resonant regime have also attracted some attention, predominantly in the case of benzene \cite{begemann2008symmetry,hettler2003current,rai2011magnetic}.
Despite theoretical progress and many experimental successes in recent years, the pursuit of novel systems exhibiting QI effects remains at the frontier of research in the field of molecular electronics \cite{stuyver2018conductance}.

\begin{figure}
    \centering
    \includegraphics{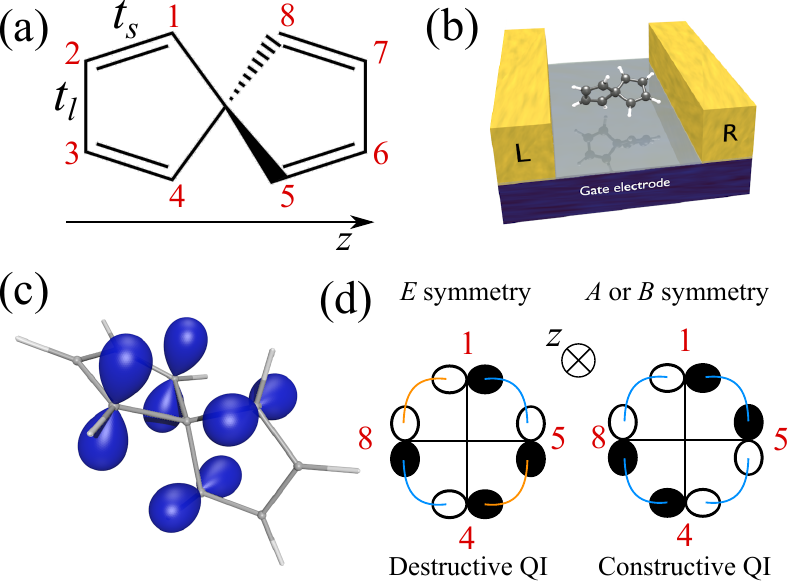}
    \caption{(a) Spiro[4.4]nonatetraene molecule. (b) Schematic of an SNT-molecular junction. (c) Schematic of the molecule showing the $p_z$ orbitals around the spiro-link. (d) The Newman projection of the spiro-link showing alignment of $p_z$ orbitals in states with $A,B$ and $E$ symmetry. }
    \label{SNT1}
\end{figure}

The focus of this work is resonant charge transport through a spiro[4.4]nonateraene (SNT) molecule, a prototypical spiro-conjugated molecular system, pictured in Fig. \ref{SNT1}. Spiro-conjugated molecules comprise conjugated moieties connected via a saturated link in such a way that the $\pi$-units are orthogonal \cite{simmons1967spiroconjugation,hoffmann1967spirarenes}. Due to this peculiar geometry, the $\pi$-units interact with each other across the spiro-link, as shown in Fig. \ref{SNT1}, giving rise to an unusual electronic structure \cite{simmons1967spiroconjugation,hoffmann1967spirarenes,batich1974equivalence}.
To the best of our knowledge these systems have not been previously studied in the single-molecule junction setting \footnote{Refs. \onlinecite{karimi2016identification,gerhard2017electrically} feature spiro-conjugated components, however, the charge transport takes place through the fully conjugated backbone.}. 

\textit{Pariser-Parr-Pople Hamiltonian.}---We wish to describe this system with a Pariser-Parr-Pople (PPP) Hamiltonian \cite{pariser1953semi,pariser1953semi2,pople1953electron}:
\begin{multline}\label{PPP}
    H_{\mathrm{PPP}} = \sum_i \varepsilon_i \; n_i + \sum_{\sigma}\sum_{<i,j>} t_{ij} (a^\dagger_{i\sigma} a_{j\sigma} + a^\dagger_{j\sigma} a_{i\sigma}) \\ + \sum_i U (n_{i \uparrow} - \frac{1}{2})(n_{i \downarrow} - \frac{1}{2})  + \sum_{i\ne j} \dfrac{V_{ij}}{2} (n_i - 1)(n_j - 1)~,
\end{multline}
where $a_{i\sigma}$ ($a_{i\sigma}^\dagger$) is the fermionic annihilation (creation) operator for an electron on site $i$ with spin $\sigma = \{\uparrow, \downarrow\}$, $n_{i \sigma} = a_{i\sigma}^\dagger a_{i\sigma}$, and $n_i =n_{i \uparrow}+ n_{i \downarrow}$.
Here, $\varepsilon_i$ is the site energy determined by the back-gate potential: $\varepsilon_i = \varepsilon_0 -e V_G$, and $U$ is the on-site electrostatic repulsion.
$V_{ij}$ is the inter-site repulsion described using the Ohno parametrisation: $V_{ij} = U/\sqrt{1 + r_{ij}^2 U^2 / 207.3 \mathrm{eV}^2}$ where $r_{ij}$ is the distance between the two $p_z$ orbitals (in \AA) \cite{ohno1964some}.
The values of the hopping integrals within each of the $\pi$ sub-units are given by: $t_{ij} = t [1 - \delta (r_{ij} - r_0)]$ where $\delta = 1.22$ \AA$^{-1}$ and $r_0 = 1.40$ \AA  \cite{bursill1998optimal,ramasesha1991optical}. The coupling across the spiro-link is described using the Hansson-Stafstr\"{o}m parametrisation (see SI for details) \cite{hansson2003intershell,hultell2007impact}.
The molecular Hamiltonian in Eq. \eqref{PPP} includes only $N=8$ sites, as numbered in Fig. \ref{SNT1}, (we ignore the $sp^3$ carbon centre \cite{hoffmann1971interaction}) -- nonetheless, as we shall show below, Eq. \eqref{PPP} can very well describe the $\pi$-electronic structure of the molecule.
Let us note here that the irreducible symmetry of the SNT molecule is $D_{2d}$, and that the system described by Eq. \eqref{PPP} is not bipartite. As a result of the latter, the Coulson--Rushbrooke pairing theorem is violated \cite{coulson1940note} and the Hamiltonian \eqref{PPP} does not exhibit electron--hole symmetry.

Since the PPP Hamiltonian is typically used only to describe fully-conjugated molecules \cite{barford2013electronic}, we begin by validating the use of $H_\mathrm{PPP}$ in the present case by considering the electronic spectrum of SNT (in the absence of the leads). We start by optimising the geometry of the molecule in Gaussian09 \cite{frisch2009gaussian} using B3LYP functional and 6-31G** basis set. This yields equilibrium inter-atomic distances, and thus determines the molecular Hamiltonian (as a function of two parameters: $t$ and $U$).  We then proceed to optimise the values of $t$ and $U$ by fitting the experimental excitation spectrum from Ref. \onlinecite{haselbach2001spiro} (solving the Hamiltonian by exact diagonalisation, see SI). The optimised parameter values, $t = - 2.36$ eV and $U = 9.31$ eV, yield a complete qualitative agreement with experimental data (in terms of the degeneracies and the singlet/triplet character of the transitions) with a relative (quantitative) error of  3.4\%. 
A graphical comparison of the theoretical and experimental values is shown in the SI. The optimised values of hopping integrals are given by $t_{15}=t_{48}=-0.24$ eV and $t_{18}=t_{45}=+0.24$ eV for coupling across the spiro-link, and $t_s = -2.52$ eV, $t_l = -2.15$ eV for coupling within the conjugated moieties, Fig. \ref{SNT1}.

\textit{Transport.}---Having validated the PPP Hamiltonian, we now proceed to the main part of this work: investigation of transport through a spiro-conjugated junction, Fig. \ref{SNT1}(b).
We assume that the molecular system is coupled to two fermionic reservoirs, the left (L) and right (R) electrode:
\begin{equation}
    H_l = \sum_{l=L,R} \sum_{k_l,\sigma} \epsilon_{k_l} c^\dagger_{k_l \sigma} c_{k_l \sigma} ~,
\end{equation}
via the Hamiltonian
\begin{equation}
    H_\mathrm{V} = \sum_{l=L,R} \sum_{k_l,\sigma} V_{k_l} a_{l\sigma}^\dagger c_{k_l \sigma} + \mathrm{H.c.} ~,
\end{equation}
where H.c.~denotes a Hermitian conjugate, and $c_{k_l\sigma}$ ($c^\dagger_{k_l\sigma}$) is the annihilation (creation) operator for an electron with energy $\epsilon_{k_l}$ and spin $\sigma$ in lead $l$. In what follows, we set $a_{L\sigma} := a_{2\sigma}$, and $a_{R\sigma} := a_{6\sigma}$ (note that all the terminal sites: 2,3,6, and 7 are equivalent by symmetry).

We continue to use the parameters obtained through the optimisation procedure described above although it should be recognised that the molecular structure can distort and the electrostatic interactions will be renormalised when a molecule is deposited in a junction (differently for every device).
Furthermore, we shall consider transport through an unfunctionalised SNT structure. Experimental studies of such a system would probably require functionalising the SNT core with the so-called anchor groups which would bind to the source and drain electrodes (although a direct connection of organic molecules to metallic electrodes is also possible \cite{kiguchi2008highly,cheng2011situ}). Nonetheless, much can be learnt by considering transport through such prototypical molecular systems as has been previously shown in both the off-resonant \cite{mitchell2017kondo,pedersen2014quantum,tsuji2016close} and resonant \cite{begemann2008symmetry,bergfield2009thermoelectric,darau2009interference} regimes.

Our focus lies in the regime of weak-molecule lead coupling, where the transport is dominated by Coulomb blockade. 
We therefore treat the electrodes perturbatively within the Born-Markov \cite{breuer2002theory} and wide-band approximation, $V_{k_l} = V_l = \mathrm{const}$. This leads to a quantum master equation for the time evolution of the reduced density matrix which is then solved in the steady-state limit, $\mathrm{d}\rho(\tau)/\mathrm{d}\tau = 0$. The effect of the leads is reduced to terms describing electron hopping on and off the molecule at the rates $\gamma_l = 2\pi \lvert V_l\rvert^2 \varrho_l$ where $\varrho_l$ is the constant density of states in the lead $l$. We use symmetric coupling throughout, $\gamma_L = \gamma_R = \gamma$. We set $\varepsilon_0 =0$, and apply the bias symmetrically: $\mu_l = \pm e V_b/2$ where $\mu_l$ is the chemical potential in the leads which (together with temperature $T$) determines the Fermi distributions, $f_l(\epsilon) = 1/(1 + \exp{[(\epsilon - \mu_l)/k_B T]})$.
\begin{figure}
    \centering
    \includegraphics{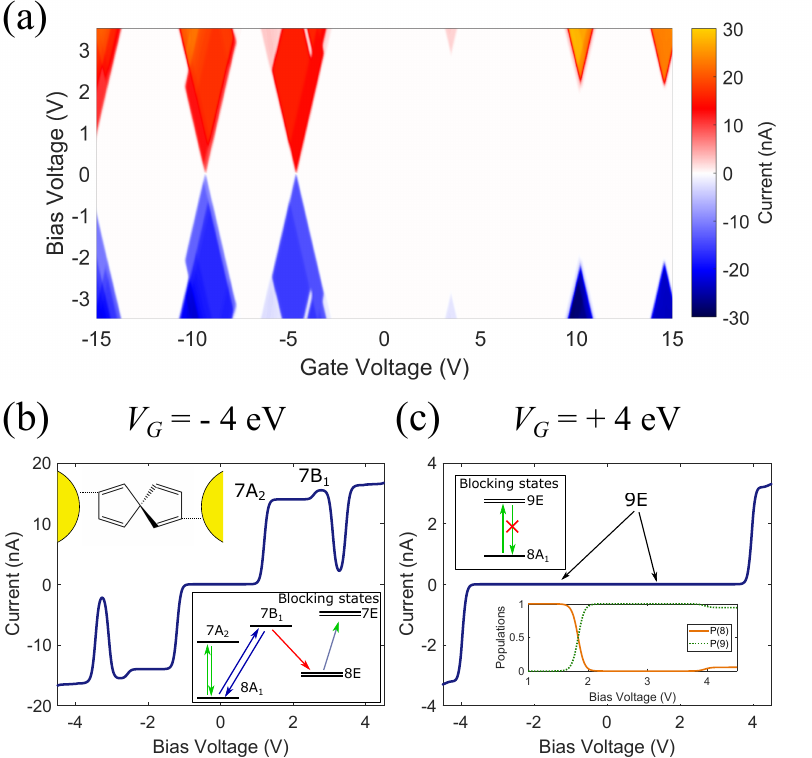}
    \caption{ (a) The stability diagram calculated for $\gamma_L = \gamma_R = 1$ meV at $T=300$ K. $IV$ characteristics at (b) $V_G = -4$ eV, (c) $V_G = +4$ eV. Parameters as in (a). The insets show transitions occurring in the regions of (b) NDC and (c) current blockade. Inset at the bottom of (c) shows populations of $N=8$ and $N=9$ charge states.}
    \label{SNT2}
\end{figure}
The size of the Fock space for our molecular Hamiltonian is $4^8$.
To make our calculation tractable we proceed to: ignore coherences between molecular states with different numbers of $\{N_\uparrow, N_\downarrow\}$ electrons (not an additional approximation) and between energy levels spaced more than $10\gamma$ apart \cite{schultz2009quantum}; and finally, consider only low-lying states (here, all states lying within at least $2eV_b$ of the ground state at a particular gate voltage).

The stability diagram (current as a function of the applied bias and gate voltage) is shown in Fig. \ref{SNT2}(a).
Two phenomena quickly become apparent: (i) Current blockade (non-closing Coulomb peaks) at positive gate voltage and (ii) Negative differential conductance (NDC, decreasing current with increasing bias voltage) present for resonant transport through various charge states.
Both phenomena can be explained by destructive quantum interference (DQI) occurring in transport through the degenerate $E$ states.
As shown in Fig. \ref{SNT1}(d), we expect DQI to occur for states with $E$ symmetry (antisymmetric with respect to the $C_2(z)$ rotation: $C_2(z) \lvert \Psi_m^N\rangle = - \lvert\Psi_m^N\rangle$), as the spiro-connected sites are coupled to each other with opposite phases.
Alternatively, one can consider the transport as occurring in the basis of eigenstates of $H_{\mathrm{PPP}}$. There, it is possible to diagonalise the molecular Hamiltonian in such a way that each of the $E$ states is localised on either moiety and therefore coupled only to a single (left or right) electrode. This corresponds to a complete destructive (inter-orbital) QI at the spiro-link. Let us stress that this is a consequence of the unusual geometry of the spiro-conjugated systems in which the interaction between the two moieties mixes only states of certain symmetry \cite{hoffmann1967spirarenes,simmons1967spiroconjugation}. 
To understand these phenomena in detail let us consider $IV$ characteristics at $V_G = \pm 4$ eV. In both cases the molecular system is found in $8 A_1$ state at $V_b = 0$ V. For negative gate voltage, as the bias is increased, the transitions to $7 A_2$ and $7 B_1$ states become possible, each of which results in a step-wise increase in current, Fig. \ref{SNT2}(b). At higher bias, the degenerate $7E$ states become populated. Since these two states interfere destructively, the transport through these states is blocked. Due to strong electron-electron repulsion \cite{xu2015negative}, further electron transfers onto the molecule cannot take place giving rise to NDC.

At $V_G = +4$ eV, the $E$ states are the lowest-lying 9-particle states. As the bias is increased, they become populated and this, again due to DQI,  results in a current blockade. In transport through the $E$ states, a complete DQI occurs at the `spiro'-link which localises the propagating charge density on either of the moieties (depending on the sign of $V_b$). 
While NDC is a feature characteristic of systems with degenerate energy levels  \cite{donarini2010interference,darau2009interference,schultz2009quantum}, the full current blockade is a consequence of the peculiar geometry of spiro-conjugated systems.
We have also investigated transport through two smaller spiro-conjugated systems. Signatures of DQI are present in both of these systems, see SI. In symmetric molecules they result from degenerate states which interfere destructively, in asymmetric ones they are a result of intra-orbital QI.

\begin{figure}
    \centering
    \includegraphics{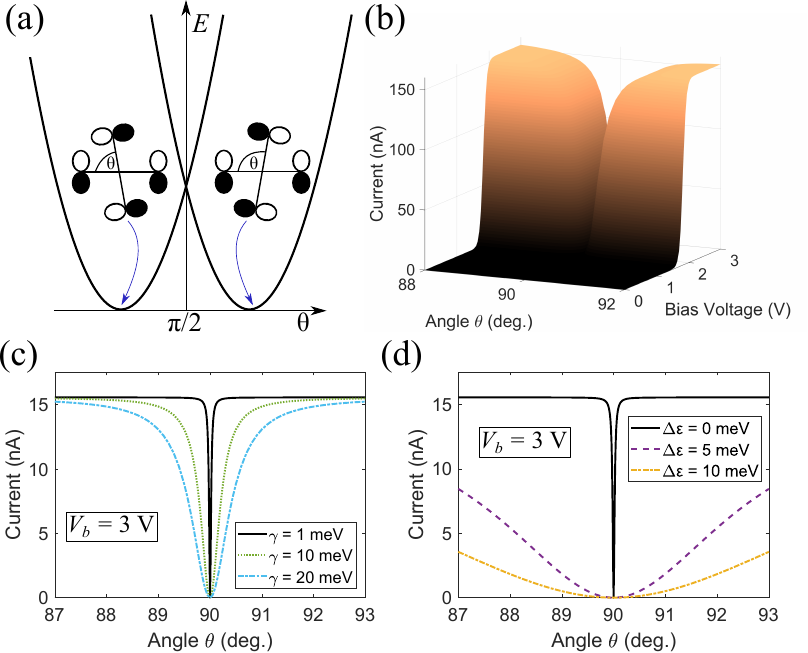}
    \caption{(a) Schematic energy diagram of the $9E$ states as a function of $\theta$. (b) $IV$ characteristics at $V_G = 4$ eV, $\gamma = 10$ meV for different values of $\theta$, . (c) Values of current at $V_G = 4$ eV, $V_b = 3$ V for different values of $\theta$ and $\gamma$; curves for $\gamma =10$ and $20$ meV were re-scaled by factors of 10 and 20 respectively. (d) Values of current at $V_G = 4$ eV, $V_b = 3$ V for different values of $\theta$ in the presence of energetic detuning between the two moieties. We shift the site energies on the two moieties by  $\pm\Delta \varepsilon$, $\gamma = 1$ meV.}
    \label{SNT3}
\end{figure}

\textit{Jahn-Teller Distortion.}---Populating degenerate electronic states, however, typically results in Jahn-Teller (JT) distortion \cite{bersuker2006jahn}. This is also the case in the SNT molecule where, upon charging, the symmetry reduces from $D_{2d}$ to $D_2$ as the two conjugated moieties undergo a twist away from the 90$^\circ$ angle ($b_1$-type distortion) \cite{haselbach2001spiro}.  As shown in Fig. \ref{SNT3}(a), upon such twisting, couplings across the spiro-link are no longer antisymmetric ($t_{15} \ne - t_{18}$ and so on). This distortion lifts the degeneracy of the 9-particle ground state (effectively reducing the overall energy of the system) and will thus have a deleterious effect on the destructive QI phenomena discussed above.
We first approach this problem from a static perspective. In Fig. \ref{SNT3}(b) we calculate the $IV$ characteristics at $V_G = +4$ eV for different values of $\theta$ - angle between the two moieties (we manually rotate one of the moieties and evaluate a new PPP Hamiltonian for each value of $\theta$).
As the angle between the two moieties is twisted away from $\pi/2$, the two orbitals cease to interfere destructively and the current blockade is lifted. Fig. \ref{SNT3}(c) is showing the values of high voltage ($V_b = 3$ V) current as a function of $\theta$. In the case of stronger molecule-lead coupling, DQI survives at significantly larger twist angles. For sufficiently strong molecule-lead coupling, Jahn-Teller distortion can be expected to have little effect on the current blockade, whereas in the case of very weak coupling the blockade is lifted even by a modest JT distortion.

It is also interesting to consider the effect of the twisting in the presence of an energetic detuning between the moieties due to capacitive coupling of the molecule to the source and drain electrodes \cite{perrin2014large,sowa2017environment}. It localises the otherwise degenerate states on either of the two moieties and thus stabilises the current blockade, see Fig. \ref{SNT3}(d). We can therefore infer that the current blockade can be stabilised by the applied bias voltage. Similar effects can be obtained by a geometric distortion of a $b_2$-type, see SI.

Microscopically, Jahn-Teller distortion originates due to coupling of the electronic $E$ states to, in this case, the twisting $b_1$ vibrational mode ($E \otimes b_1$ type) \cite{bersuker2006jahn,haselbach2001spiro}. We proceed to examine the interplay between JT distortion and DQI within this microscopic picture.
In what follows we will consider only 5 electronic states: the totally symmetric 8-particle ground state and four (spin and spatially) degenerate $9E$ states (two spatially degenerate levels are denoted as $\alpha$ and $\beta$; all the other states lie outside the bias window for $V_b$ and $V_G$ considered henceforth). The  $\alpha$ and $\beta$ states do not vary with the displacement coordinate (although, crucially, their energies do) and so we define these two states as obtained through the diagonalisation of the Hamiltonian at $\theta = \pi/2 + \delta\theta$ where $\delta\theta \rightarrow 0$.
The relevant Hamiltonian can now be written as \cite{bersuker2006jahn,schultz2010quantum}:
\begin{multline} \label{H_JT}
 H_{\mathrm{JT}} = \varepsilon' (n_\alpha + n_\beta) + \omega b^\dagger b + g (b^\dagger + b) (n_\alpha - n_\beta)\\ +\sum_{l, k_l,\sigma} \epsilon_{k_l} c^\dagger_{k_l \sigma} c_{k_l \sigma} + \sum_{l, k_l, \sigma} V_{l} d_{l\sigma}^\dagger c_{k_l \sigma} + \mathrm{H.c.} ~,  
\end{multline}
where $n_\alpha = \sum_\sigma n_{\alpha \sigma}$, $\varepsilon'$ is the energy difference between the $8A_1$ and $9E$ states, $g$ is the electron-phonon coupling constant, and $\omega$ is the frequency of the $b_1$ mode in question with raising (lowering) operator $b^\dagger$ ($b$). The operator $d_{l\sigma}^\dagger$ ($d_{l\sigma}$) describes (de-)charging of the molecule at contact $l$ and can be written as $d_{l\sigma}^\dagger = \zeta_{l \alpha} a^\dagger_{\alpha \sigma} + \zeta_{l \beta} a^\dagger_{\beta \sigma}$ where the coefficients $\zeta_{l \alpha}$ and $\zeta_{l \beta }$ are obtained using the PPP Hamiltonian.

\begin{figure}
    \centering
    \includegraphics{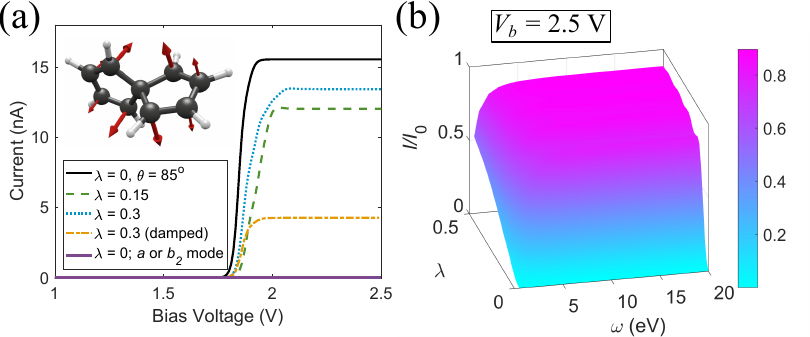}
    \caption{(a) $IV$ characteristics for different values of the electron-phonon coupling parameter $\lambda$ for $\omega = 27$ meV and hopping rate: $\gamma = 1$ meV. The damping rate is taken to be $\gamma_{d} = 0.1$ meV where appropriate, $T = 77$ K. (b) Values of current at $V_b = 2.5$ V as a function of $\lambda$ and $\omega$ renormalised by $I_0$ - value of current for $\lambda = 0$ and $\theta =85^\circ$. The molecule-lead rate is $\gamma = 20$ meV, $T = 300$ K. 50 vibrational levels were included in both calculations.}
    \label{SNT4}
\end{figure}
We proceed to describe this problem within the Born-Markov approximation \cite{breuer2002theory} with respect to the leads in the polaron-transformed frame while allowing for non-equilibrium dynamics  of the vibrational mode (unless specified otherwise), and use standard methods to extract the current from the resulting quantum master equation \cite{flindt2004full,hartle2009vibrational,sowa2017vibrational}.
In Fig. \ref{SNT4}(a), we plot the $IV$ characteristics obtained for different values of the dimensionless electron-phonon coupling parameter, $\lambda = g/\omega$. 
Firstly, we can observe that coupling to the mode of $b_1$ symmetry lifts the current blockade with current increasing with $\lambda$. The latter can be contrasted with what occurs typically in a simple single-level case where the inclusion of vibrational interactions leads to a suppression of current \cite{koch2005franck}. While full non-equilibrium dynamics of the $b_1$ mode can rather efficiently lift the blockade, even very slow damping (introduced phenomenologically \cite{breuer2002theory}) can significantly reduce the current flowing through the system.
In Fig. \ref{SNT4}(b) we consider values of current at high bias as a function of $\omega$ and $\lambda$. Once again, we see that increasing $\lambda$ yields higher values of current.
The fact that current decreases with decreasing $\omega$ can be explained as follows. Within the Franck-Condon principle, the electron hopping on and off the molecule occurs within a rigid nuclear framework. For $\omega \ll \gamma$, when the unitary evolution of the mode is much slower then the (de-)charging of the molecule, this unitary evolution of the vibrational mode will become the bottleneck of the overall electron transport. 

In contrast, distortions induced by $a_1$ and $a_2$ vibrational modes do not split the $9E$ states, and are therefore incapable of lifting the current blockade. 
Coupling to vibrational modes of $b_2$ symmetry similarly cannot do that despite also resulting in Jahn-Teller distortion (to $C_{2v}$ symmetry). This situation can be described by a Hamiltonian equivalent to \eqref{H_JT} but with (redefined) $\alpha$ and $\beta$ localised on either of the two conjugated moieties.
Including coupling to both $b_1$ and $b_2$ modes would require inclusion of non-adiabatic effects which are beyond the scope of this work. We can anticipate, however, that they will stabilise the current blockade, similarly to what is inferred from the static model.

\textit{Conclusions.}---We have studied resonant transport through a prototypical spiro-conjugated single-molecule junction which we have described with an (experimentally validated) PPP Hamiltonian. We have shown that in the regime of strong electron-electron repulsion, destructive QI effects lead to current blockade and NDC. While NDC is characteristic for systems with doubly-degenerate levels, the current blockade of the type discussed here results from the peculiar structure of spiro-conjugated systems.
The aforementioned effects result from destructive interference between the degenerate $E$ states. Populating these states will, however, result in Jahn-Teller distortion which may lift the DQI-induced phenomena.
We predict that the signatures of QI can nonetheless survive, especially for a relatively strong molecule-lead coupling, and slow nuclear dynamics.
The strength of the molecule-lead coupling can be controlled to some extent by the choice of molecule-lead coupling or even by mechanical means \cite{frisenda2016transition}, and the frequency of the vibrational modes can be strongly influenced by chemical design, see SI.
Let us also note that functionalising the molecular core can result in symmetry breaking and result in small splittings of the $E$ states (\textit{cf.} Refs. \onlinecite{ballmann2012experimental, hartle2011quantum}) with the pseudo-Jahn-Teller distortion taking place instead \cite{bersuker2006jahn}.

Jahn-Teller distortion is a ubiquitous phenomenon present throughout chemistry and condensed matter physics. As shown herein, spiro-conjugated systems can serve as an excellent platform for studying its effects in the transport setting.
Finally, we have shown that destructive quantum interference effects are inherently present in transport through spiro-conjugated systems which may thus serve as useful building blocks in molecular thermoelectrics, transistors or rectifiers.

\begin{acknowledgments}
The authors thank Colin Lambert, William Barford, James Thomas and Roald Hoffmann for useful discussions.
J.K.S. thanks the Clarendon Fund, Hertford College and EPSRC for financial support. E.M.G. acknowledges funding from the Royal Society of Edinburgh and the Scottish Government, J.A.M. acknowledges funding from the Royal Academy of Engineering. We also thank the John Tempelton Foundation and acknowledge the use of the University of Oxford Advanced Research Computing (ARC) facility in carrying out this work (http://dx.doi.org/10.5281/zenodo.22558).
\end{acknowledgments}


\begin{thebibliography}{56}%
	\makeatletter
	\providecommand \@ifxundefined [1]{%
		\@ifx{#1\undefined}
	}%
	\providecommand \@ifnum [1]{%
		\ifnum #1\expandafter \@firstoftwo
		\else \expandafter \@secondoftwo
		\fi
	}%
	\providecommand \@ifx [1]{%
		\ifx #1\expandafter \@firstoftwo
		\else \expandafter \@secondoftwo
		\fi
	}%
	\providecommand \natexlab [1]{#1}%
	\providecommand \enquote  [1]{``#1''}%
	\providecommand \bibnamefont  [1]{#1}%
	\providecommand \bibfnamefont [1]{#1}%
	\providecommand \citenamefont [1]{#1}%
	\providecommand \href@noop [0]{\@secondoftwo}%
	\providecommand \href [0]{\begingroup \@sanitize@url \@href}%
	\providecommand \@href[1]{\@@startlink{#1}\@@href}%
	\providecommand \@@href[1]{\endgroup#1\@@endlink}%
	\providecommand \@sanitize@url [0]{\catcode `\\12\catcode `\$12\catcode
		`\&12\catcode `\#12\catcode `\^12\catcode `\_12\catcode `\%12\relax}%
	\providecommand \@@startlink[1]{}%
	\providecommand \@@endlink[0]{}%
	\providecommand \url  [0]{\begingroup\@sanitize@url \@url }%
	\providecommand \@url [1]{\endgroup\@href {#1}{\urlprefix }}%
	\providecommand \urlprefix  [0]{URL }%
	\providecommand \Eprint [0]{\href }%
	\providecommand \doibase [0]{http://dx.doi.org/}%
	\providecommand \selectlanguage [0]{\@gobble}%
	\providecommand \bibinfo  [0]{\@secondoftwo}%
	\providecommand \bibfield  [0]{\@secondoftwo}%
	\providecommand \translation [1]{[#1]}%
	\providecommand \BibitemOpen [0]{}%
	\providecommand \bibitemStop [0]{}%
	\providecommand \bibitemNoStop [0]{.\EOS\space}%
	\providecommand \EOS [0]{\spacefactor3000\relax}%
	\providecommand \BibitemShut  [1]{\csname bibitem#1\endcsname}%
	\let\auto@bib@innerbib\@empty
	\bibitem [{\citenamefont {Walter}\ \emph {et~al.}(2004)\citenamefont {Walter},
		\citenamefont {Neuhauser},\ and\ \citenamefont {Baer}}]{walter2004quantum}%
	\BibitemOpen
	\bibfield  {author} {\bibinfo {author} {\bibfnamefont {D.}~\bibnamefont
			{Walter}}, \bibinfo {author} {\bibfnamefont {D.}~\bibnamefont {Neuhauser}}, \
		and\ \bibinfo {author} {\bibfnamefont {R.}~\bibnamefont {Baer}},\ }\href@noop
	{} {\bibfield  {journal} {\bibinfo  {journal} {Chem. Phys.}\ }\textbf
		{\bibinfo {volume} {299}},\ \bibinfo {pages} {139} (\bibinfo {year}
		{2004})}\BibitemShut {NoStop}%
	\bibitem [{\citenamefont {Ke}\ \emph {et~al.}(2008)\citenamefont {Ke},
		\citenamefont {Yang},\ and\ \citenamefont {Baranger}}]{ke2008quantum}%
	\BibitemOpen
	\bibfield  {author} {\bibinfo {author} {\bibfnamefont {S.-H.}\ \bibnamefont
			{Ke}}, \bibinfo {author} {\bibfnamefont {W.}~\bibnamefont {Yang}}, \ and\
		\bibinfo {author} {\bibfnamefont {H.~U.}\ \bibnamefont {Baranger}},\
	}\href@noop {} {\bibfield  {journal} {\bibinfo  {journal} {Nano Lett.}\
	}\textbf {\bibinfo {volume} {8}},\ \bibinfo {pages} {3257} (\bibinfo {year}
	{2008})}\BibitemShut {NoStop}%
\bibitem [{\citenamefont {Cardamone}\ \emph {et~al.}(2006)\citenamefont
	{Cardamone}, \citenamefont {Stafford},\ and\ \citenamefont
	{Mazumdar}}]{cardamone2006controlling}%
\BibitemOpen
\bibfield  {author} {\bibinfo {author} {\bibfnamefont {D.~M.}\ \bibnamefont
		{Cardamone}}, \bibinfo {author} {\bibfnamefont {C.~A.}\ \bibnamefont
		{Stafford}}, \ and\ \bibinfo {author} {\bibfnamefont {S.}~\bibnamefont
		{Mazumdar}},\ }\href@noop {} {\bibfield  {journal} {\bibinfo  {journal} {Nano
			Lett.}\ }\textbf {\bibinfo {volume} {6}},\ \bibinfo {pages} {2422} (\bibinfo
	{year} {2006})}\BibitemShut {NoStop}%
\bibitem [{\citenamefont {Arroyo}\ \emph {et~al.}(2013)\citenamefont {Arroyo},
	\citenamefont {Tarkuc}, \citenamefont {Frisenda}, \citenamefont
	{Seldenthuis}, \citenamefont {Woerde}, \citenamefont {Eelkema}, \citenamefont
	{Grozema},\ and\ \citenamefont {van~der Zant}}]{arroyo2013signatures}%
\BibitemOpen
\bibfield  {author} {\bibinfo {author} {\bibfnamefont {C.~R.}\ \bibnamefont
		{Arroyo}}, \bibinfo {author} {\bibfnamefont {S.}~\bibnamefont {Tarkuc}},
	\bibinfo {author} {\bibfnamefont {R.}~\bibnamefont {Frisenda}}, \bibinfo
	{author} {\bibfnamefont {J.~S.}\ \bibnamefont {Seldenthuis}}, \bibinfo
	{author} {\bibfnamefont {C.~H.~M.}\ \bibnamefont {Woerde}}, \bibinfo {author}
	{\bibfnamefont {R.}~\bibnamefont {Eelkema}}, \bibinfo {author} {\bibfnamefont
		{F.~C.}\ \bibnamefont {Grozema}}, \ and\ \bibinfo {author} {\bibfnamefont
		{H.~S.~J.}\ \bibnamefont {van~der Zant}},\ }\href@noop {} {\bibfield
	{journal} {\bibinfo  {journal} {Angew. Chem.}\ }\textbf {\bibinfo {volume}
		{125}},\ \bibinfo {pages} {3234} (\bibinfo {year} {2013})}\BibitemShut
{NoStop}%
\bibitem [{\citenamefont {Pickup}\ \emph {et~al.}(2015)\citenamefont {Pickup},
	\citenamefont {Fowler}, \citenamefont {Borg},\ and\ \citenamefont
	{Sciriha}}]{pickup2015new}%
\BibitemOpen
\bibfield  {author} {\bibinfo {author} {\bibfnamefont {B.~T.}\ \bibnamefont
		{Pickup}}, \bibinfo {author} {\bibfnamefont {P.~W.}\ \bibnamefont {Fowler}},
	\bibinfo {author} {\bibfnamefont {M.}~\bibnamefont {Borg}}, \ and\ \bibinfo
	{author} {\bibfnamefont {I.}~\bibnamefont {Sciriha}},\ }\href@noop {}
{\bibfield  {journal} {\bibinfo  {journal} {J. Chem. Phys.}\ }\textbf
	{\bibinfo {volume} {143}},\ \bibinfo {pages} {194105} (\bibinfo {year}
	{2015})}\BibitemShut {NoStop}%
\bibitem [{\citenamefont {Solomon}\ \emph
	{et~al.}(2008{\natexlab{a}})\citenamefont {Solomon}, \citenamefont {Andrews},
	\citenamefont {Goldsmith}, \citenamefont {Hansen}, \citenamefont
	{Wasielewski}, \citenamefont {Van~Duyne},\ and\ \citenamefont
	{Ratner}}]{solomon2008quantum}%
\BibitemOpen
\bibfield  {author} {\bibinfo {author} {\bibfnamefont {G.~C.}\ \bibnamefont
		{Solomon}}, \bibinfo {author} {\bibfnamefont {D.~Q.}\ \bibnamefont
		{Andrews}}, \bibinfo {author} {\bibfnamefont {R.~H.}\ \bibnamefont
		{Goldsmith}}, \bibinfo {author} {\bibfnamefont {T.}~\bibnamefont {Hansen}},
	\bibinfo {author} {\bibfnamefont {M.~R.}\ \bibnamefont {Wasielewski}},
	\bibinfo {author} {\bibfnamefont {R.~P.}\ \bibnamefont {Van~Duyne}}, \ and\
	\bibinfo {author} {\bibfnamefont {M.~A.}\ \bibnamefont {Ratner}},\
}\href@noop {} {\bibfield  {journal} {\bibinfo  {journal} {J. Am. Chem.
		Soc.}\ }\textbf {\bibinfo {volume} {130}},\ \bibinfo {pages} {17301}
(\bibinfo {year} {2008}{\natexlab{a}})}\BibitemShut {NoStop}%
\bibitem [{\citenamefont {Gu{\'e}don}\ \emph {et~al.}(2012)\citenamefont
	{Gu{\'e}don}, \citenamefont {Valkenier}, \citenamefont {Markussen},
	\citenamefont {Thygesen}, \citenamefont {Hummelen},\ and\ \citenamefont {Van
		Der~Molen}}]{guedon2012observation}%
\BibitemOpen
\bibfield  {author} {\bibinfo {author} {\bibfnamefont {C.~M.}\ \bibnamefont
		{Gu{\'e}don}}, \bibinfo {author} {\bibfnamefont {H.}~\bibnamefont
		{Valkenier}}, \bibinfo {author} {\bibfnamefont {T.}~\bibnamefont
		{Markussen}}, \bibinfo {author} {\bibfnamefont {K.~S.}\ \bibnamefont
		{Thygesen}}, \bibinfo {author} {\bibfnamefont {J.~C.}\ \bibnamefont
		{Hummelen}}, \ and\ \bibinfo {author} {\bibfnamefont {S.~J.}\ \bibnamefont
		{Van Der~Molen}},\ }\href@noop {} {\bibfield  {journal} {\bibinfo  {journal}
		{Nat. Nanotechnol.}\ }\textbf {\bibinfo {volume} {7}},\ \bibinfo {pages}
	{305} (\bibinfo {year} {2012})}\BibitemShut {NoStop}%
\bibitem [{\citenamefont {Solomon}\ \emph
	{et~al.}(2008{\natexlab{b}})\citenamefont {Solomon}, \citenamefont {Andrews},
	\citenamefont {Van~Duyne},\ and\ \citenamefont {Ratner}}]{solomon2008things}%
\BibitemOpen
\bibfield  {author} {\bibinfo {author} {\bibfnamefont {G.~C.}\ \bibnamefont
		{Solomon}}, \bibinfo {author} {\bibfnamefont {D.~Q.}\ \bibnamefont
		{Andrews}}, \bibinfo {author} {\bibfnamefont {R.~P.}\ \bibnamefont
		{Van~Duyne}}, \ and\ \bibinfo {author} {\bibfnamefont {M.~A.}\ \bibnamefont
		{Ratner}},\ }\href@noop {} {\bibfield  {journal} {\bibinfo  {journal} {J. Am.
			Chem. Soc.}\ }\textbf {\bibinfo {volume} {130}},\ \bibinfo {pages} {7788}
	(\bibinfo {year} {2008}{\natexlab{b}})}\BibitemShut {NoStop}%
\bibitem [{\citenamefont {Koole}\ \emph {et~al.}(2015)\citenamefont {Koole},
	\citenamefont {Thijssen}, \citenamefont {Valkenier}, \citenamefont
	{Hummelen},\ and\ \citenamefont {van~der Zant}}]{koole2015electric}%
\BibitemOpen
\bibfield  {author} {\bibinfo {author} {\bibfnamefont {M.}~\bibnamefont
		{Koole}}, \bibinfo {author} {\bibfnamefont {J.~M.}\ \bibnamefont {Thijssen}},
	\bibinfo {author} {\bibfnamefont {H.}~\bibnamefont {Valkenier}}, \bibinfo
	{author} {\bibfnamefont {J.~C.}\ \bibnamefont {Hummelen}}, \ and\ \bibinfo
	{author} {\bibfnamefont {H.~S.~J.}\ \bibnamefont {van~der Zant}},\
}\href@noop {} {\bibfield  {journal} {\bibinfo  {journal} {Nano Lett.}\
}\textbf {\bibinfo {volume} {15}},\ \bibinfo {pages} {5569} (\bibinfo {year}
{2015})}\BibitemShut {NoStop}%
\bibitem [{\citenamefont {Lambert}(2015)}]{lambert2015basic}%
\BibitemOpen
\bibfield  {author} {\bibinfo {author} {\bibfnamefont {C.~J.}\ \bibnamefont
		{Lambert}},\ }\href@noop {} {\bibfield  {journal} {\bibinfo  {journal} {Chem.
			Soc. Rev.}\ }\textbf {\bibinfo {volume} {44}},\ \bibinfo {pages} {875}
	(\bibinfo {year} {2015})}\BibitemShut {NoStop}%
\bibitem [{\citenamefont {Papadopoulos}\ \emph {et~al.}(2006)\citenamefont
	{Papadopoulos}, \citenamefont {Grace},\ and\ \citenamefont
	{Lambert}}]{papadopoulos2006control}%
\BibitemOpen
\bibfield  {author} {\bibinfo {author} {\bibfnamefont {T.~A.}\ \bibnamefont
		{Papadopoulos}}, \bibinfo {author} {\bibfnamefont {I.~M.}\ \bibnamefont
		{Grace}}, \ and\ \bibinfo {author} {\bibfnamefont {C.~J.}\ \bibnamefont
		{Lambert}},\ }\href@noop {} {\bibfield  {journal} {\bibinfo  {journal} {Phys.
			Rev. B}\ }\textbf {\bibinfo {volume} {74}},\ \bibinfo {pages} {193306}
	(\bibinfo {year} {2006})}\BibitemShut {NoStop}%
\bibitem [{\citenamefont {Begemann}\ \emph {et~al.}(2008)\citenamefont
	{Begemann}, \citenamefont {Darau}, \citenamefont {Donarini},\ and\
	\citenamefont {Grifoni}}]{begemann2008symmetry}%
\BibitemOpen
\bibfield  {author} {\bibinfo {author} {\bibfnamefont {G.}~\bibnamefont
		{Begemann}}, \bibinfo {author} {\bibfnamefont {D.}~\bibnamefont {Darau}},
	\bibinfo {author} {\bibfnamefont {A.}~\bibnamefont {Donarini}}, \ and\
	\bibinfo {author} {\bibfnamefont {M.}~\bibnamefont {Grifoni}},\ }\href@noop
{} {\bibfield  {journal} {\bibinfo  {journal} {Phys. Rev. B}\ }\textbf
	{\bibinfo {volume} {77}},\ \bibinfo {pages} {201406} (\bibinfo {year}
	{2008})}\BibitemShut {NoStop}%
\bibitem [{\citenamefont {Hettler}\ \emph {et~al.}(2003)\citenamefont
	{Hettler}, \citenamefont {Wenzel}, \citenamefont {Wegewijs},\ and\
	\citenamefont {Schoeller}}]{hettler2003current}%
\BibitemOpen
\bibfield  {author} {\bibinfo {author} {\bibfnamefont {M.~H.}\ \bibnamefont
		{Hettler}}, \bibinfo {author} {\bibfnamefont {W.}~\bibnamefont {Wenzel}},
	\bibinfo {author} {\bibfnamefont {M.~R.}\ \bibnamefont {Wegewijs}}, \ and\
	\bibinfo {author} {\bibfnamefont {H.}~\bibnamefont {Schoeller}},\ }\href@noop
{} {\bibfield  {journal} {\bibinfo  {journal} {Phys. Rev. Lett.}\ }\textbf
	{\bibinfo {volume} {90}},\ \bibinfo {pages} {076805} (\bibinfo {year}
	{2003})}\BibitemShut {NoStop}%
\bibitem [{\citenamefont {Rai}\ \emph {et~al.}(2011)\citenamefont {Rai},
	\citenamefont {Hod},\ and\ \citenamefont {Nitzan}}]{rai2011magnetic}%
\BibitemOpen
\bibfield  {author} {\bibinfo {author} {\bibfnamefont {D.}~\bibnamefont
		{Rai}}, \bibinfo {author} {\bibfnamefont {O.}~\bibnamefont {Hod}}, \ and\
	\bibinfo {author} {\bibfnamefont {A.}~\bibnamefont {Nitzan}},\ }\href@noop {}
{\bibfield  {journal} {\bibinfo  {journal} {J. Phys. Chem. Lett.}\ }\textbf
	{\bibinfo {volume} {2}},\ \bibinfo {pages} {2118} (\bibinfo {year}
	{2011})}\BibitemShut {NoStop}%
\bibitem [{\citenamefont {Stuyver}\ \emph {et~al.}(2018)\citenamefont
	{Stuyver}, \citenamefont {Perrin}, \citenamefont {Geerlings}, \citenamefont
	{De~Proft},\ and\ \citenamefont {Alonso}}]{stuyver2018conductance}%
\BibitemOpen
\bibfield  {author} {\bibinfo {author} {\bibfnamefont {T.}~\bibnamefont
		{Stuyver}}, \bibinfo {author} {\bibfnamefont {M.~L.}\ \bibnamefont {Perrin}},
	\bibinfo {author} {\bibfnamefont {P.}~\bibnamefont {Geerlings}}, \bibinfo
	{author} {\bibfnamefont {F.}~\bibnamefont {De~Proft}}, \ and\ \bibinfo
	{author} {\bibfnamefont {M.}~\bibnamefont {Alonso}},\ }\href@noop {}
{\bibfield  {journal} {\bibinfo  {journal} {J. Am. Chem. Soc.}\ }\textbf
	{\bibinfo {volume} {140}},\ \bibinfo {pages} {1313} (\bibinfo {year}
	{2018})}\BibitemShut {NoStop}%
\bibitem [{\citenamefont {Simmons}\ and\ \citenamefont
	{Fukunaga}(1967)}]{simmons1967spiroconjugation}%
\BibitemOpen
\bibfield  {author} {\bibinfo {author} {\bibfnamefont {H.~E.}\ \bibnamefont
		{Simmons}}\ and\ \bibinfo {author} {\bibfnamefont {T.}~\bibnamefont
		{Fukunaga}},\ }\href@noop {} {\bibfield  {journal} {\bibinfo  {journal} {J.
			Am. Chem. Soc.}\ }\textbf {\bibinfo {volume} {89}},\ \bibinfo {pages} {5208}
	(\bibinfo {year} {1967})}\BibitemShut {NoStop}%
\bibitem [{\citenamefont {Hoffmann}\ \emph {et~al.}(1967)\citenamefont
	{Hoffmann}, \citenamefont {Imamura},\ and\ \citenamefont
	{Zeiss}}]{hoffmann1967spirarenes}%
\BibitemOpen
\bibfield  {author} {\bibinfo {author} {\bibfnamefont {R.}~\bibnamefont
		{Hoffmann}}, \bibinfo {author} {\bibfnamefont {A.}~\bibnamefont {Imamura}}, \
	and\ \bibinfo {author} {\bibfnamefont {G.~D.}\ \bibnamefont {Zeiss}},\
}\href@noop {} {\bibfield  {journal} {\bibinfo  {journal} {J. Am. Chem.
		Soc.}\ }\textbf {\bibinfo {volume} {89}},\ \bibinfo {pages} {5215} (\bibinfo
{year} {1967})}\BibitemShut {NoStop}%
\bibitem [{\citenamefont {Batich}\ \emph {et~al.}(1974)\citenamefont {Batich},
	\citenamefont {Heilbronner}, \citenamefont {Rommel}, \citenamefont
	{Semmelhack},\ and\ \citenamefont {Foos}}]{batich1974equivalence}%
\BibitemOpen
\bibfield  {author} {\bibinfo {author} {\bibfnamefont {C.}~\bibnamefont
		{Batich}}, \bibinfo {author} {\bibfnamefont {E.}~\bibnamefont {Heilbronner}},
	\bibinfo {author} {\bibfnamefont {E.}~\bibnamefont {Rommel}}, \bibinfo
	{author} {\bibfnamefont {M.~F.}\ \bibnamefont {Semmelhack}}, \ and\ \bibinfo
	{author} {\bibfnamefont {J.~S.}\ \bibnamefont {Foos}},\ }\href@noop {}
{\bibfield  {journal} {\bibinfo  {journal} {J. Am. Chem. Soc.}\ }\textbf
	{\bibinfo {volume} {96}},\ \bibinfo {pages} {7662} (\bibinfo {year}
	{1974})}\BibitemShut {NoStop}%
\bibitem [{Note1()}]{Note1}%
\BibitemOpen
\bibinfo {note} {Refs. \protect \rev@citealp
	{karimi2016identification,gerhard2017electrically} feature spiro-conjugated
	components, however, the charge transport takes place through the fully
	conjugated backbone.}\BibitemShut {Stop}%
\bibitem [{\citenamefont {Pariser}\ and\ \citenamefont
	{Parr}(1953{\natexlab{a}})}]{pariser1953semi}%
\BibitemOpen
\bibfield  {author} {\bibinfo {author} {\bibfnamefont {R.}~\bibnamefont
		{Pariser}}\ and\ \bibinfo {author} {\bibfnamefont {R.~G.}\ \bibnamefont
		{Parr}},\ }\href@noop {} {\bibfield  {journal} {\bibinfo  {journal} {J. Chem.
			Phys.}\ }\textbf {\bibinfo {volume} {21}},\ \bibinfo {pages} {767} (\bibinfo
	{year} {1953}{\natexlab{a}})}\BibitemShut {NoStop}%
\bibitem [{\citenamefont {Pariser}\ and\ \citenamefont
	{Parr}(1953{\natexlab{b}})}]{pariser1953semi2}%
\BibitemOpen
\bibfield  {author} {\bibinfo {author} {\bibfnamefont {R.}~\bibnamefont
		{Pariser}}\ and\ \bibinfo {author} {\bibfnamefont {R.~G.}\ \bibnamefont
		{Parr}},\ }\href@noop {} {\bibfield  {journal} {\bibinfo  {journal} {J. Chem.
			Phys.}\ }\textbf {\bibinfo {volume} {21}},\ \bibinfo {pages} {466} (\bibinfo
	{year} {1953}{\natexlab{b}})}\BibitemShut {NoStop}%
\bibitem [{\citenamefont {Pople}(1953)}]{pople1953electron}%
\BibitemOpen
\bibfield  {author} {\bibinfo {author} {\bibfnamefont {J.~A.}\ \bibnamefont
		{Pople}},\ }\href@noop {} {\bibfield  {journal} {\bibinfo  {journal} {Trans.
			Faraday Soc.}\ }\textbf {\bibinfo {volume} {49}},\ \bibinfo {pages} {1375}
	(\bibinfo {year} {1953})}\BibitemShut {NoStop}%
\bibitem [{\citenamefont {Ohno}(1964)}]{ohno1964some}%
\BibitemOpen
\bibfield  {author} {\bibinfo {author} {\bibfnamefont {K.}~\bibnamefont
		{Ohno}},\ }\href@noop {} {\bibfield  {journal} {\bibinfo  {journal} {Theor.
			Chim. Acta}\ }\textbf {\bibinfo {volume} {2}},\ \bibinfo {pages} {219}
	(\bibinfo {year} {1964})}\BibitemShut {NoStop}%
\bibitem [{\citenamefont {Bursill}\ \emph {et~al.}(1998)\citenamefont
	{Bursill}, \citenamefont {Castleton},\ and\ \citenamefont
	{Barford}}]{bursill1998optimal}%
\BibitemOpen
\bibfield  {author} {\bibinfo {author} {\bibfnamefont {R.~J.}\ \bibnamefont
		{Bursill}}, \bibinfo {author} {\bibfnamefont {C.}~\bibnamefont {Castleton}},
	\ and\ \bibinfo {author} {\bibfnamefont {W.}~\bibnamefont {Barford}},\
}\href@noop {} {\bibfield  {journal} {\bibinfo  {journal} {Chem. Phys.
		Lett.}\ }\textbf {\bibinfo {volume} {294}},\ \bibinfo {pages} {305} (\bibinfo
{year} {1998})}\BibitemShut {NoStop}%
\bibitem [{\citenamefont {Ramasesha}\ \emph {et~al.}(1991)\citenamefont
	{Ramasesha}, \citenamefont {Albert},\ and\ \citenamefont
	{Sinha}}]{ramasesha1991optical}%
\BibitemOpen
\bibfield  {author} {\bibinfo {author} {\bibfnamefont {S.}~\bibnamefont
		{Ramasesha}}, \bibinfo {author} {\bibfnamefont {I.~D.~L.}\ \bibnamefont
		{Albert}}, \ and\ \bibinfo {author} {\bibfnamefont {B.}~\bibnamefont
		{Sinha}},\ }\href@noop {} {\bibfield  {journal} {\bibinfo  {journal} {Mol.
			Phys.}\ }\textbf {\bibinfo {volume} {72}},\ \bibinfo {pages} {537} (\bibinfo
	{year} {1991})}\BibitemShut {NoStop}%
\bibitem [{\citenamefont {Hansson}\ and\ \citenamefont
	{Stafstr{\"o}m}(2003)}]{hansson2003intershell}%
\BibitemOpen
\bibfield  {author} {\bibinfo {author} {\bibfnamefont {A.}~\bibnamefont
		{Hansson}}\ and\ \bibinfo {author} {\bibfnamefont {S.}~\bibnamefont
		{Stafstr{\"o}m}},\ }\href@noop {} {\bibfield  {journal} {\bibinfo  {journal}
		{Phys. Rev. B}\ }\textbf {\bibinfo {volume} {67}},\ \bibinfo {pages} {075406}
	(\bibinfo {year} {2003})}\BibitemShut {NoStop}%
\bibitem [{\citenamefont {Hultell}\ and\ \citenamefont
	{Stafstr{\"o}m}(2007)}]{hultell2007impact}%
\BibitemOpen
\bibfield  {author} {\bibinfo {author} {\bibfnamefont {M.}~\bibnamefont
		{Hultell}}\ and\ \bibinfo {author} {\bibfnamefont {S.}~\bibnamefont
		{Stafstr{\"o}m}},\ }\href@noop {} {\bibfield  {journal} {\bibinfo  {journal}
		{Phys. Rev. B}\ }\textbf {\bibinfo {volume} {75}},\ \bibinfo {pages} {104304}
	(\bibinfo {year} {2007})}\BibitemShut {NoStop}%
\bibitem [{\citenamefont {Hoffmann}(1971)}]{hoffmann1971interaction}%
\BibitemOpen
\bibfield  {author} {\bibinfo {author} {\bibfnamefont {R.}~\bibnamefont
		{Hoffmann}},\ }\href@noop {} {\bibfield  {journal} {\bibinfo  {journal} {Acc.
			Chem. Res.}\ }\textbf {\bibinfo {volume} {4}},\ \bibinfo {pages} {1}
	(\bibinfo {year} {1971})}\BibitemShut {NoStop}%
\bibitem [{\citenamefont {Coulson}\ and\ \citenamefont
	{Rushbrooke}(1940)}]{coulson1940note}%
\BibitemOpen
\bibfield  {author} {\bibinfo {author} {\bibfnamefont {C.~A.}\ \bibnamefont
		{Coulson}}\ and\ \bibinfo {author} {\bibfnamefont {G.~S.}\ \bibnamefont
		{Rushbrooke}},\ }\href@noop {} {\bibfield  {journal} {\bibinfo  {journal}
		{Math. Proc. Camb. Phil. Soc.}\ }\textbf {\bibinfo {volume} {36}},\ \bibinfo
	{pages} {193} (\bibinfo {year} {1940})}\BibitemShut {NoStop}%
\bibitem [{\citenamefont {Barford}(2013)}]{barford2013electronic}%
\BibitemOpen
\bibfield  {author} {\bibinfo {author} {\bibfnamefont {W.}~\bibnamefont
		{Barford}},\ }\href@noop {} {\emph {\bibinfo {title} {Electronic and optical
			properties of conjugated polymers}}},\ Vol.\ \bibinfo {volume} {159}\
(\bibinfo  {publisher} {Oxford University Press},\ \bibinfo {year}
{2013})\BibitemShut {NoStop}%
\bibitem [{\citenamefont {Frisch}\ \emph {et~al.}(2009)\citenamefont {Frisch},
	\citenamefont {Trucks}, \citenamefont {Schlegel}, \citenamefont {Scuseria},
	\citenamefont {Robb}, \citenamefont {Cheeseman}, \citenamefont {Scalmani},
	\citenamefont {Barone}, \citenamefont {Mennucci}, \citenamefont {Petersson}
	\emph {et~al.}}]{frisch2009gaussian}%
\BibitemOpen
\bibfield  {author} {\bibinfo {author} {\bibfnamefont {M.~J.}\ \bibnamefont
		{Frisch}}, \bibinfo {author} {\bibfnamefont {G.~W.}\ \bibnamefont {Trucks}},
	\bibinfo {author} {\bibfnamefont {H.~B.}\ \bibnamefont {Schlegel}}, \bibinfo
	{author} {\bibfnamefont {G.~E.}\ \bibnamefont {Scuseria}}, \bibinfo {author}
	{\bibfnamefont {M.~A.}\ \bibnamefont {Robb}}, \bibinfo {author}
	{\bibfnamefont {J.~R.}\ \bibnamefont {Cheeseman}}, \bibinfo {author}
	{\bibfnamefont {G.}~\bibnamefont {Scalmani}}, \bibinfo {author}
	{\bibfnamefont {V.}~\bibnamefont {Barone}}, \bibinfo {author} {\bibfnamefont
		{B.}~\bibnamefont {Mennucci}}, \bibinfo {author} {\bibfnamefont {G.~A.}\
		\bibnamefont {Petersson}},  \emph {et~al.},\ }\href@noop {} {\bibfield
	{journal} {\bibinfo  {journal} {Inc., Wallingford, CT}\ }\textbf {\bibinfo
		{volume} {200}} (\bibinfo {year} {2009})}\BibitemShut {NoStop}%
\bibitem [{\citenamefont {Haselbach}\ \emph {et~al.}(2001)\citenamefont
	{Haselbach}, \citenamefont {Allan}, \citenamefont {Bally}, \citenamefont
	{Bednarek}, \citenamefont {Sergenton}, \citenamefont {De~Meijere},
	\citenamefont {Kozhushkov}, \citenamefont {Piacenza},\ and\ \citenamefont
	{Grimme}}]{haselbach2001spiro}%
\BibitemOpen
\bibfield  {author} {\bibinfo {author} {\bibfnamefont {E.}~\bibnamefont
		{Haselbach}}, \bibinfo {author} {\bibfnamefont {M.}~\bibnamefont {Allan}},
	\bibinfo {author} {\bibfnamefont {T.}~\bibnamefont {Bally}}, \bibinfo
	{author} {\bibfnamefont {P.}~\bibnamefont {Bednarek}}, \bibinfo {author}
	{\bibfnamefont {A.-C.}\ \bibnamefont {Sergenton}}, \bibinfo {author}
	{\bibfnamefont {A.}~\bibnamefont {De~Meijere}}, \bibinfo {author}
	{\bibfnamefont {S.}~\bibnamefont {Kozhushkov}}, \bibinfo {author}
	{\bibfnamefont {M.}~\bibnamefont {Piacenza}}, \ and\ \bibinfo {author}
	{\bibfnamefont {S.}~\bibnamefont {Grimme}},\ }\href@noop {} {\bibfield
	{journal} {\bibinfo  {journal} {Helv. Chim. Acta}\ }\textbf {\bibinfo
		{volume} {84}},\ \bibinfo {pages} {1670} (\bibinfo {year}
	{2001})}\BibitemShut {NoStop}%
\bibitem [{\citenamefont {Kiguchi}\ \emph {et~al.}(2008)\citenamefont
	{Kiguchi}, \citenamefont {Tal}, \citenamefont {Wohlthat}, \citenamefont
	{Pauly}, \citenamefont {Krieger}, \citenamefont {Djukic}, \citenamefont
	{Cuevas},\ and\ \citenamefont {van Ruitenbeek}}]{kiguchi2008highly}%
\BibitemOpen
\bibfield  {author} {\bibinfo {author} {\bibfnamefont {M.}~\bibnamefont
		{Kiguchi}}, \bibinfo {author} {\bibfnamefont {O.}~\bibnamefont {Tal}},
	\bibinfo {author} {\bibfnamefont {S.}~\bibnamefont {Wohlthat}}, \bibinfo
	{author} {\bibfnamefont {F.}~\bibnamefont {Pauly}}, \bibinfo {author}
	{\bibfnamefont {M.}~\bibnamefont {Krieger}}, \bibinfo {author} {\bibfnamefont
		{D.}~\bibnamefont {Djukic}}, \bibinfo {author} {\bibfnamefont {J.~C.}\
		\bibnamefont {Cuevas}}, \ and\ \bibinfo {author} {\bibfnamefont {J.~M.}\
		\bibnamefont {van Ruitenbeek}},\ }\href@noop {} {\bibfield  {journal}
	{\bibinfo  {journal} {Phys. Rev. Lett.}\ }\textbf {\bibinfo {volume} {101}},\
	\bibinfo {pages} {046801} (\bibinfo {year} {2008})}\BibitemShut {NoStop}%
\bibitem [{\citenamefont {Cheng}\ \emph {et~al.}(2011)\citenamefont {Cheng},
	\citenamefont {Skouta}, \citenamefont {Vazquez}, \citenamefont {Widawsky},
	\citenamefont {Schneebeli}, \citenamefont {Chen}, \citenamefont {Hybertsen},
	\citenamefont {Breslow},\ and\ \citenamefont {Venkataraman}}]{cheng2011situ}%
\BibitemOpen
\bibfield  {author} {\bibinfo {author} {\bibfnamefont {Z.-L.}\ \bibnamefont
		{Cheng}}, \bibinfo {author} {\bibfnamefont {R.}~\bibnamefont {Skouta}},
	\bibinfo {author} {\bibfnamefont {H.}~\bibnamefont {Vazquez}}, \bibinfo
	{author} {\bibfnamefont {J.~R.}\ \bibnamefont {Widawsky}}, \bibinfo {author}
	{\bibfnamefont {S.}~\bibnamefont {Schneebeli}}, \bibinfo {author}
	{\bibfnamefont {W.}~\bibnamefont {Chen}}, \bibinfo {author} {\bibfnamefont
		{M.~S.}\ \bibnamefont {Hybertsen}}, \bibinfo {author} {\bibfnamefont
		{R.}~\bibnamefont {Breslow}}, \ and\ \bibinfo {author} {\bibfnamefont
		{L.}~\bibnamefont {Venkataraman}},\ }\href@noop {} {\bibfield  {journal}
	{\bibinfo  {journal} {Nat. Nanotechnol.}\ }\textbf {\bibinfo {volume} {6}},\
	\bibinfo {pages} {353} (\bibinfo {year} {2011})}\BibitemShut {NoStop}%
\bibitem [{\citenamefont {Mitchell}\ \emph {et~al.}(2017)\citenamefont
	{Mitchell}, \citenamefont {Pedersen}, \citenamefont {Hedeg{\aa}rd},\ and\
	\citenamefont {Paaske}}]{mitchell2017kondo}%
\BibitemOpen
\bibfield  {author} {\bibinfo {author} {\bibfnamefont {A.~K.}\ \bibnamefont
		{Mitchell}}, \bibinfo {author} {\bibfnamefont {K.~G.~L.}\ \bibnamefont
		{Pedersen}}, \bibinfo {author} {\bibfnamefont {P.}~\bibnamefont
		{Hedeg{\aa}rd}}, \ and\ \bibinfo {author} {\bibfnamefont {J.}~\bibnamefont
		{Paaske}},\ }\href@noop {} {\bibfield  {journal} {\bibinfo  {journal} {Nat.
			Commun.}\ }\textbf {\bibinfo {volume} {8}} (\bibinfo {year}
	{2017})}\BibitemShut {NoStop}%
\bibitem [{\citenamefont {Pedersen}\ \emph {et~al.}(2014)\citenamefont
	{Pedersen}, \citenamefont {Strange}, \citenamefont {Leijnse}, \citenamefont
	{Hedeg{\aa}rd}, \citenamefont {Solomon},\ and\ \citenamefont
	{Paaske}}]{pedersen2014quantum}%
\BibitemOpen
\bibfield  {author} {\bibinfo {author} {\bibfnamefont {K.~G.~L.}\
		\bibnamefont {Pedersen}}, \bibinfo {author} {\bibfnamefont {M.}~\bibnamefont
		{Strange}}, \bibinfo {author} {\bibfnamefont {M.}~\bibnamefont {Leijnse}},
	\bibinfo {author} {\bibfnamefont {P.}~\bibnamefont {Hedeg{\aa}rd}}, \bibinfo
	{author} {\bibfnamefont {G.~C.}\ \bibnamefont {Solomon}}, \ and\ \bibinfo
	{author} {\bibfnamefont {J.}~\bibnamefont {Paaske}},\ }\href@noop {}
{\bibfield  {journal} {\bibinfo  {journal} {Phys. Rev. B}\ }\textbf {\bibinfo
		{volume} {90}},\ \bibinfo {pages} {125413} (\bibinfo {year}
	{2014})}\BibitemShut {NoStop}%
\bibitem [{\citenamefont {Tsuji}\ \emph {et~al.}(2016)\citenamefont {Tsuji},
	\citenamefont {Hoffmann}, \citenamefont {Strange},\ and\ \citenamefont
	{Solomon}}]{tsuji2016close}%
\BibitemOpen
\bibfield  {author} {\bibinfo {author} {\bibfnamefont {Y.}~\bibnamefont
		{Tsuji}}, \bibinfo {author} {\bibfnamefont {R.}~\bibnamefont {Hoffmann}},
	\bibinfo {author} {\bibfnamefont {M.}~\bibnamefont {Strange}}, \ and\
	\bibinfo {author} {\bibfnamefont {G.~C.}\ \bibnamefont {Solomon}},\
}\href@noop {} {\bibfield  {journal} {\bibinfo  {journal} {Proc. Natl. Acad.
		Sci. U.S.A.}\ }\textbf {\bibinfo {volume} {113}},\ \bibinfo {pages} {E413}
(\bibinfo {year} {2016})}\BibitemShut {NoStop}%
\bibitem [{\citenamefont {Bergfield}\ and\ \citenamefont
	{Stafford}(2009)}]{bergfield2009thermoelectric}%
\BibitemOpen
\bibfield  {author} {\bibinfo {author} {\bibfnamefont {J.~P.}\ \bibnamefont
		{Bergfield}}\ and\ \bibinfo {author} {\bibfnamefont {C.~A.}\ \bibnamefont
		{Stafford}},\ }\href@noop {} {\bibfield  {journal} {\bibinfo  {journal} {Nano
			Lett.}\ }\textbf {\bibinfo {volume} {9}},\ \bibinfo {pages} {3072} (\bibinfo
	{year} {2009})}\BibitemShut {NoStop}%
\bibitem [{\citenamefont {Darau}\ \emph {et~al.}(2009)\citenamefont {Darau},
	\citenamefont {Begemann}, \citenamefont {Donarini},\ and\ \citenamefont
	{Grifoni}}]{darau2009interference}%
\BibitemOpen
\bibfield  {author} {\bibinfo {author} {\bibfnamefont {D.}~\bibnamefont
		{Darau}}, \bibinfo {author} {\bibfnamefont {G.}~\bibnamefont {Begemann}},
	\bibinfo {author} {\bibfnamefont {A.}~\bibnamefont {Donarini}}, \ and\
	\bibinfo {author} {\bibfnamefont {M.}~\bibnamefont {Grifoni}},\ }\href@noop
{} {\bibfield  {journal} {\bibinfo  {journal} {Phys. Rev. B}\ }\textbf
	{\bibinfo {volume} {79}},\ \bibinfo {pages} {235404} (\bibinfo {year}
	{2009})}\BibitemShut {NoStop}%
\bibitem [{\citenamefont {Breuer}\ and\ \citenamefont
	{Petruccione}(2002)}]{breuer2002theory}%
\BibitemOpen
\bibfield  {author} {\bibinfo {author} {\bibfnamefont {H.-P.}\ \bibnamefont
		{Breuer}}\ and\ \bibinfo {author} {\bibfnamefont {F.}~\bibnamefont
		{Petruccione}},\ }\href@noop {} {\emph {\bibinfo {title} {The Theory of Open
			Quantum Systems}}}\ (\bibinfo  {publisher} {Oxford University Press},\
\bibinfo {year} {2002})\BibitemShut {NoStop}%
\bibitem [{\citenamefont {Schultz}\ and\ \citenamefont {von
		Oppen}(2009)}]{schultz2009quantum}%
\BibitemOpen
\bibfield  {author} {\bibinfo {author} {\bibfnamefont {M.~G.}\ \bibnamefont
		{Schultz}}\ and\ \bibinfo {author} {\bibfnamefont {F.}~\bibnamefont {von
			Oppen}},\ }\href@noop {} {\bibfield  {journal} {\bibinfo  {journal} {Phys.
			Rev. B}\ }\textbf {\bibinfo {volume} {80}},\ \bibinfo {pages} {033302}
	(\bibinfo {year} {2009})}\BibitemShut {NoStop}%
\bibitem [{\citenamefont {Xu}\ and\ \citenamefont
	{Dubi}(2015)}]{xu2015negative}%
\BibitemOpen
\bibfield  {author} {\bibinfo {author} {\bibfnamefont {B.}~\bibnamefont
		{Xu}}\ and\ \bibinfo {author} {\bibfnamefont {Y.}~\bibnamefont {Dubi}},\
}\href@noop {} {\bibfield  {journal} {\bibinfo  {journal} {J. Phys.: Condens.
		Matter}\ }\textbf {\bibinfo {volume} {27}},\ \bibinfo {pages} {263202}
(\bibinfo {year} {2015})}\BibitemShut {NoStop}%
\bibitem [{\citenamefont {Donarini}\ \emph {et~al.}(2010)\citenamefont
	{Donarini}, \citenamefont {Begemann},\ and\ \citenamefont
	{Grifoni}}]{donarini2010interference}%
\BibitemOpen
\bibfield  {author} {\bibinfo {author} {\bibfnamefont {A.}~\bibnamefont
		{Donarini}}, \bibinfo {author} {\bibfnamefont {G.}~\bibnamefont {Begemann}},
	\ and\ \bibinfo {author} {\bibfnamefont {M.}~\bibnamefont {Grifoni}},\
}\href@noop {} {\bibfield  {journal} {\bibinfo  {journal} {Phys. Rev. B}\
}\textbf {\bibinfo {volume} {82}},\ \bibinfo {pages} {125451} (\bibinfo
{year} {2010})}\BibitemShut {NoStop}%
\bibitem [{\citenamefont {Bersuker}(2006)}]{bersuker2006jahn}%
\BibitemOpen
\bibfield  {author} {\bibinfo {author} {\bibfnamefont {I.~B.}\ \bibnamefont
		{Bersuker}},\ }\href@noop {} {\emph {\bibinfo {title} {The Jahn-Teller
			Effect}}}\ (\bibinfo  {publisher} {Cambridge University Press},\ \bibinfo
{year} {2006})\BibitemShut {NoStop}%
\bibitem [{\citenamefont {Perrin}\ \emph {et~al.}(2014)\citenamefont {Perrin},
	\citenamefont {Frisenda}, \citenamefont {Koole}, \citenamefont {Seldenthuis},
	\citenamefont {Gil}, \citenamefont {Valkenier}, \citenamefont {Hummelen},
	\citenamefont {Renaud}, \citenamefont {Grozema}, \citenamefont {Thijssen}
	\emph {et~al.}}]{perrin2014large}%
\BibitemOpen
\bibfield  {author} {\bibinfo {author} {\bibfnamefont {M.~L.}\ \bibnamefont
		{Perrin}}, \bibinfo {author} {\bibfnamefont {R.}~\bibnamefont {Frisenda}},
	\bibinfo {author} {\bibfnamefont {M.}~\bibnamefont {Koole}}, \bibinfo
	{author} {\bibfnamefont {J.~S.}\ \bibnamefont {Seldenthuis}}, \bibinfo
	{author} {\bibfnamefont {J.~A.~C.}\ \bibnamefont {Gil}}, \bibinfo {author}
	{\bibfnamefont {H.}~\bibnamefont {Valkenier}}, \bibinfo {author}
	{\bibfnamefont {J.~C.}\ \bibnamefont {Hummelen}}, \bibinfo {author}
	{\bibfnamefont {N.}~\bibnamefont {Renaud}}, \bibinfo {author} {\bibfnamefont
		{F.~C.}\ \bibnamefont {Grozema}}, \bibinfo {author} {\bibfnamefont {J.~M.}\
		\bibnamefont {Thijssen}},  \emph {et~al.},\ }\href@noop {} {\bibfield
	{journal} {\bibinfo  {journal} {Nat. Nanotechnol.}\ }\textbf {\bibinfo
		{volume} {9}},\ \bibinfo {pages} {830} (\bibinfo {year} {2014})}\BibitemShut
{NoStop}%
\bibitem [{\citenamefont {Sowa}\ \emph
	{et~al.}(2017{\natexlab{a}})\citenamefont {Sowa}, \citenamefont {Mol},
	\citenamefont {Briggs},\ and\ \citenamefont {Gauger}}]{sowa2017environment}%
\BibitemOpen
\bibfield  {author} {\bibinfo {author} {\bibfnamefont {J.~K.}\ \bibnamefont
		{Sowa}}, \bibinfo {author} {\bibfnamefont {J.~A.}\ \bibnamefont {Mol}},
	\bibinfo {author} {\bibfnamefont {G.~A.~D.}\ \bibnamefont {Briggs}}, \ and\
	\bibinfo {author} {\bibfnamefont {E.~M.}\ \bibnamefont {Gauger}},\
}\href@noop {} {\bibfield  {journal} {\bibinfo  {journal} {Phys. Chem. Chem.
		Phys.}\ } (\bibinfo {year} {2017}{\natexlab{a}})}\BibitemShut {NoStop}%
\bibitem [{\citenamefont {Schultz}(2010)}]{schultz2010quantum}%
\BibitemOpen
\bibfield  {author} {\bibinfo {author} {\bibfnamefont {M.~G.}\ \bibnamefont
		{Schultz}},\ }\href@noop {} {\bibfield  {journal} {\bibinfo  {journal} {Phys.
			Rev. B}\ }\textbf {\bibinfo {volume} {82}},\ \bibinfo {pages} {155408}
	(\bibinfo {year} {2010})}\BibitemShut {NoStop}%
\bibitem [{\citenamefont {Flindt}\ \emph {et~al.}(2004)\citenamefont {Flindt},
	\citenamefont {Novotn{\`y}},\ and\ \citenamefont {Jauho}}]{flindt2004full}%
\BibitemOpen
\bibfield  {author} {\bibinfo {author} {\bibfnamefont {C.}~\bibnamefont
		{Flindt}}, \bibinfo {author} {\bibfnamefont {T.}~\bibnamefont {Novotn{\`y}}},
	\ and\ \bibinfo {author} {\bibfnamefont {A.-P.}\ \bibnamefont {Jauho}},\
}\href@noop {} {\bibfield  {journal} {\bibinfo  {journal} {EPL}\ }\textbf
{\bibinfo {volume} {69}},\ \bibinfo {pages} {475} (\bibinfo {year}
{2004})}\BibitemShut {NoStop}%
\bibitem [{\citenamefont {H{\"a}rtle}\ \emph {et~al.}(2009)\citenamefont
	{H{\"a}rtle}, \citenamefont {Benesch},\ and\ \citenamefont
	{Thoss}}]{hartle2009vibrational}%
\BibitemOpen
\bibfield  {author} {\bibinfo {author} {\bibfnamefont {R.}~\bibnamefont
		{H{\"a}rtle}}, \bibinfo {author} {\bibfnamefont {C.}~\bibnamefont {Benesch}},
	\ and\ \bibinfo {author} {\bibfnamefont {M.}~\bibnamefont {Thoss}},\
}\href@noop {} {\bibfield  {journal} {\bibinfo  {journal} {Phys. Rev. Lett.}\
}\textbf {\bibinfo {volume} {102}},\ \bibinfo {pages} {146801} (\bibinfo
{year} {2009})}\BibitemShut {NoStop}%
\bibitem [{\citenamefont {Sowa}\ \emph
	{et~al.}(2017{\natexlab{b}})\citenamefont {Sowa}, \citenamefont {Mol},
	\citenamefont {Briggs},\ and\ \citenamefont {Gauger}}]{sowa2017vibrational}%
\BibitemOpen
\bibfield  {author} {\bibinfo {author} {\bibfnamefont {J.~K.}\ \bibnamefont
		{Sowa}}, \bibinfo {author} {\bibfnamefont {J.~A.}\ \bibnamefont {Mol}},
	\bibinfo {author} {\bibfnamefont {G.~A.~D.}\ \bibnamefont {Briggs}}, \ and\
	\bibinfo {author} {\bibfnamefont {E.~M.}\ \bibnamefont {Gauger}},\
}\href@noop {} {\bibfield  {journal} {\bibinfo  {journal} {Phys. Rev. B}\
}\textbf {\bibinfo {volume} {95}},\ \bibinfo {pages} {085423} (\bibinfo
{year} {2017}{\natexlab{b}})}\BibitemShut {NoStop}%
\bibitem [{\citenamefont {Koch}\ and\ \citenamefont {von
		Oppen}(2005)}]{koch2005franck}%
\BibitemOpen
\bibfield  {author} {\bibinfo {author} {\bibfnamefont {J.}~\bibnamefont
		{Koch}}\ and\ \bibinfo {author} {\bibfnamefont {F.}~\bibnamefont {von
			Oppen}},\ }\href@noop {} {\bibfield  {journal} {\bibinfo  {journal} {Phys.
			Rev. Lett.}\ }\textbf {\bibinfo {volume} {94}},\ \bibinfo {pages} {206804}
	(\bibinfo {year} {2005})}\BibitemShut {NoStop}%
\bibitem [{\citenamefont {Frisenda}\ and\ \citenamefont {van~der
		Zant}(2016)}]{frisenda2016transition}%
\BibitemOpen
\bibfield  {author} {\bibinfo {author} {\bibfnamefont {R.}~\bibnamefont
		{Frisenda}}\ and\ \bibinfo {author} {\bibfnamefont {H.~S.~J.}\ \bibnamefont
		{van~der Zant}},\ }\href@noop {} {\bibfield  {journal} {\bibinfo  {journal}
		{Phys. Rev. Lett.}\ }\textbf {\bibinfo {volume} {117}},\ \bibinfo {pages}
	{126804} (\bibinfo {year} {2016})}\BibitemShut {NoStop}%
\bibitem [{\citenamefont {Ballmann}\ \emph {et~al.}(2012)\citenamefont
	{Ballmann}, \citenamefont {H{\"a}rtle}, \citenamefont {Coto}, \citenamefont
	{Elbing}, \citenamefont {Mayor}, \citenamefont {Bryce}, \citenamefont
	{Thoss},\ and\ \citenamefont {Weber}}]{ballmann2012experimental}%
\BibitemOpen
\bibfield  {author} {\bibinfo {author} {\bibfnamefont {S.}~\bibnamefont
		{Ballmann}}, \bibinfo {author} {\bibfnamefont {R.}~\bibnamefont
		{H{\"a}rtle}}, \bibinfo {author} {\bibfnamefont {P.~B.}\ \bibnamefont
		{Coto}}, \bibinfo {author} {\bibfnamefont {M.}~\bibnamefont {Elbing}},
	\bibinfo {author} {\bibfnamefont {M.}~\bibnamefont {Mayor}}, \bibinfo
	{author} {\bibfnamefont {M.~R.}\ \bibnamefont {Bryce}}, \bibinfo {author}
	{\bibfnamefont {M.}~\bibnamefont {Thoss}}, \ and\ \bibinfo {author}
	{\bibfnamefont {H.~B.}\ \bibnamefont {Weber}},\ }\href@noop {} {\bibfield
	{journal} {\bibinfo  {journal} {Phys. Rev. Lett.}\ }\textbf {\bibinfo
		{volume} {109}},\ \bibinfo {pages} {056801} (\bibinfo {year}
	{2012})}\BibitemShut {NoStop}%
\bibitem [{\citenamefont {H{\"a}rtle}\ \emph {et~al.}(2011)\citenamefont
	{H{\"a}rtle}, \citenamefont {Butzin}, \citenamefont {Rubio-Pons},\ and\
	\citenamefont {Thoss}}]{hartle2011quantum}%
\BibitemOpen
\bibfield  {author} {\bibinfo {author} {\bibfnamefont {R.}~\bibnamefont
		{H{\"a}rtle}}, \bibinfo {author} {\bibfnamefont {M.}~\bibnamefont {Butzin}},
	\bibinfo {author} {\bibfnamefont {O.}~\bibnamefont {Rubio-Pons}}, \ and\
	\bibinfo {author} {\bibfnamefont {M.}~\bibnamefont {Thoss}},\ }\href@noop {}
{\bibfield  {journal} {\bibinfo  {journal} {Phys. Rev. Lett.}\ }\textbf
	{\bibinfo {volume} {107}},\ \bibinfo {pages} {046802} (\bibinfo {year}
	{2011})}\BibitemShut {NoStop}%
\bibitem [{\citenamefont {Karimi}\ \emph {et~al.}(2016)\citenamefont {Karimi},
	\citenamefont {Bahoosh}, \citenamefont {Val{\'a}{\v{s}}ek}, \citenamefont
	{B{\"u}rkle}, \citenamefont {Mayor}, \citenamefont {Pauly},\ and\
	\citenamefont {Scheer}}]{karimi2016identification}%
\BibitemOpen
\bibfield  {author} {\bibinfo {author} {\bibfnamefont {M.~A.}\ \bibnamefont
		{Karimi}}, \bibinfo {author} {\bibfnamefont {S.~G.}\ \bibnamefont {Bahoosh}},
	\bibinfo {author} {\bibfnamefont {M.}~\bibnamefont {Val{\'a}{\v{s}}ek}},
	\bibinfo {author} {\bibfnamefont {M.}~\bibnamefont {B{\"u}rkle}}, \bibinfo
	{author} {\bibfnamefont {M.}~\bibnamefont {Mayor}}, \bibinfo {author}
	{\bibfnamefont {F.}~\bibnamefont {Pauly}}, \ and\ \bibinfo {author}
	{\bibfnamefont {E.}~\bibnamefont {Scheer}},\ }\href@noop {} {\bibfield
	{journal} {\bibinfo  {journal} {Nanoscale}\ }\textbf {\bibinfo {volume}
		{8}},\ \bibinfo {pages} {10582} (\bibinfo {year} {2016})}\BibitemShut
{NoStop}%
\bibitem [{\citenamefont {Gerhard}\ \emph {et~al.}(2017)\citenamefont
	{Gerhard}, \citenamefont {Edelmann}, \citenamefont {Homberg}, \citenamefont
	{Val{\'a}{\v{s}}ek}, \citenamefont {Bahoosh}, \citenamefont {Lukas},
	\citenamefont {Pauly}, \citenamefont {Mayor},\ and\ \citenamefont
	{Wulfhekel}}]{gerhard2017electrically}%
\BibitemOpen
\bibfield  {author} {\bibinfo {author} {\bibfnamefont {L.}~\bibnamefont
		{Gerhard}}, \bibinfo {author} {\bibfnamefont {K.}~\bibnamefont {Edelmann}},
	\bibinfo {author} {\bibfnamefont {J.}~\bibnamefont {Homberg}}, \bibinfo
	{author} {\bibfnamefont {M.}~\bibnamefont {Val{\'a}{\v{s}}ek}}, \bibinfo
	{author} {\bibfnamefont {S.~G.}\ \bibnamefont {Bahoosh}}, \bibinfo {author}
	{\bibfnamefont {M.}~\bibnamefont {Lukas}}, \bibinfo {author} {\bibfnamefont
		{F.}~\bibnamefont {Pauly}}, \bibinfo {author} {\bibfnamefont
		{M.}~\bibnamefont {Mayor}}, \ and\ \bibinfo {author} {\bibfnamefont
		{W.}~\bibnamefont {Wulfhekel}},\ }\href@noop {} {\bibfield  {journal}
	{\bibinfo  {journal} {Nat. Commun.}\ }\textbf {\bibinfo {volume} {8}},\
	\bibinfo {pages} {14672} (\bibinfo {year} {2017})}\BibitemShut {NoStop}%
\end{thebibliography}
\end{document}